\documentstyle[aps,eqsecnum]{revtex}
\begin{document}
\widetext
\def\qed{\hfill \vrule height 7pt width 7pt depth 0pt \smallskip}
\renewcommand{\pl}{\partial}
\newcommand{\ra}{\rightarrow} \newcommand{\al}{\alpha} \newcommand{\st}{\sqrt}
\newcommand{\hf}{\frac{1}{2}} \newcommand{\iy}{\infty} \newcommand{\la}{\lambda}
\newcommand{\bq}{\begin{equation}} \newcommand{\eq}{\end{equation}}
\newcommand{\qr}{\frac{1}{4}} 
\newcommand{\fr}{\frac}
\def\up#1{\leavevmode \raise .3ex\hbox{$#1$}}
\def\down#1{\leavevmode \lower .5ex\hbox{$\scriptstyle#1$}}
\def\chJ{\up{\chi}_{\down{J}}}
\def\sdown#1{\leavevmode \lower .2ex\hbox{$\scriptstyle#1$}}
\title{
Level Spacing Distributions and the Bessel Kernel}
\author{Craig A.~Tracy\footnotemark[1]}
\footnotetext[1]{e-mail address: tracy@itd.ucdavis.edu}
\address{
Department of Mathematics and Institute of Theoretical Dynamics,\\
University of California,
Davis, CA 95616, USA}
\author{Harold Widom\footnotemark[2]}
\footnotetext[2]{e-mail address: widom@cats.ucsc.edu}
\address{
Department of Mathematics,\\
University of California,
Santa Cruz, CA 95064, USA}
\maketitle
\begin{abstract}
Scaling models of random $N\times N$ hermitian matrices and passing to the
limit $N\ra\iy$ leads to integral operators whose Fredholm determinants
describe the statistics of the spacing of the eigenvalues of hermitian
matrices of large order.
For the  Gaussian Unitary Ensemble, and for many others
as well,
the kernel one obtains by scaling in the
``bulk'' of the spectrum is the ``sine kernel''
$ \frac{\sin\pi(x-y)}{\pi(x-y)}$.
Rescaling the GUE at the ``edge'' of the spectrum leads to the kernel
$\frac{{\rm Ai}(x)\,{\rm Ai}'(y)\,-{\rm Ai}'(x)\,{\rm Ai}(y)}{x-y}$
where ${\rm Ai}$ is the Airy function.
In  previous work 
we found several
analogies between properties of this ``Airy kernel'' and known properties
 of the sine
kernel:
 a system of partial differential equations associated with the logarithmic
differential
 of the Fredholm determinant when the underlying domain is a union of
intervals; a representation
 of the Fredholm determinant in terms of a 
Painlev{\'e}
transcendent in the case of a single interval; and, also in this case,
asymptotic
expansions for these determinants and related quantities,
 achieved with the help of
a differential operator which commutes with the integral operator.
In this paper
 we show that there are completely analogous properties for a class of
kernels which arise when one rescales
 the Laguerre or Jacobi ensembles at the edge
of the spectrum, namely
\[\frac{ J_{\al}(\sqrt{x})\,\sqrt{y} J_{\al}'(\sqrt{y})\,-
\sqrt{x} J_{\al}'(\sqrt{x})\, J_{\al}(\sqrt{y})}{2(x-y)}\]
where $J_\al(z)$ is the Bessel function of order $\al$.
In the cases $\al=\mp\hf$ these become, after a variable change,
the
kernels which arise when taking scaling limits in the bulk of the spectrum
for the Gaussian orthogonal and symplectic ensembles. In
particular, an asymptotic expansion we derive will generalize ones found by
Dyson for the Fredholm determinants of these kernels.
\end{abstract}

\section{Introduction and Statement of Results}
\subsection{Introduction}
Scaling models of random $N\times N$ hermitian matrices and passing to the
limit $N\ra\iy$ leads to integral operators whose Fredholm determinants
describe the statistics of the spacing of the eigenvalues of hermitian 
matrices of large order \cite{mehta_book,porter}.
 Which integral operators (or, more precisely,
which kernels of integral operators) result depends on the matrix model 
one starts with and at which location in the spectrum the scaling takes
place.

For the simplest model, the Gaussian Unitary Ensemble (GUE), and for many others
as well (see, e.g., \cite{kpw,mahoux_mehta91,nagao_wadati,pastur}),
the kernel one obtains by scaling in the
``bulk'' of the spectrum is the ``sine kernel''
\[\frac{\sin\pi(x-y)}{\pi(x-y)}.\]
Precisely, this comes about as follows. If $\{\phi_{k}(x)\}_{k=0}^{\iy}$ is the sequence
obtained by orthonormalizing  the sequence $\{x^{k}\,e^{-x^{2}/2}\}$ over $(-\iy,\iy)$ and
if 
\bq K_{N}(x,y)\,=\,\sum_{k=0}^{N-1}\,\phi_{k}(x)\,\phi_{k}(y)\, ,  \label{KN} \eq
then in the GUE the probability density that $n$ of the eigenvalues (irrespective of 
order) lie in infinitesimal intervals about $x_{1},\ldots,x_{n}$ is equal to
\[ R_{n}(x_{1},\ldots,x_{n})\,=\,\mbox{det}\ (K_{N}(x_{i},x_{j}))\,_{i,j=1,\ldots,n}.\]
The density of eigenvalues at a fixed point $z$ is $R_{1}(z)$, and this is 
$\sim\sqrt{2N}/\pi$ as $N\ra\iy$. Rescaling at $z$ leads to the sine kernel because
of the relation
\[\lim_{N\ra\iy}\frac{\pi}{\sqrt{2N}}K_{N}(z+\frac{\pi x}{\sqrt{2N}},z+\frac{\pi y}
{\sqrt{2N}})\,=\,\frac{\sin\pi(x-y)}{\pi(x-y)}.\]

Rescaling the GUE at the ``edge'' of the spectrum, however, leads to a different
kernel. The edge corresponds to $z\sim\sqrt{2N}$, at which point the density is 
$\sim2^{\hf}N^{\frac{1}{6}}$, and we have there the scaling limit 
\cite{bowick_brezin,forrester,moore}
\[\lim_{N\ra\iy}\frac{1}{2^{\hf}N^{\frac{1}{6}}} K_{N}(\sqrt{2N}+
\frac{x}{2^{\hf}N^{\frac{1}{6}}},\sqrt{2N}+\frac{y}{2^{\hf}N^{\frac{1}{6}}})\,
=\,\frac{{\rm Ai}(x)\,{\rm Ai}'(y)\,-{\rm Ai}'(x)\,{\rm Ai}(y)}{x-y}\]
where ${\rm Ai}$ is the Airy function. In  previous work \cite{tw2} 
we found several
analogies between properties of this ``Airy kernel''  and known properties
 of the sine
kernel: a system of partial differential equations associated with the logarithmic
differential of the Fredholm determinant when the underlying domain is a union of
intervals \cite{jmms}; a representation of the Fredholm determinant in terms of a Painlev{\'e}
transcendent in the case of a single interval \cite{jmms}; and, also in this case, asymptotic
expansions for these determinants and related quantities
\cite{dyson76,btw,widom92,dyson92,mehta_mahoux92b}, achieved with the help of
a differential operator which commutes with the integral operator. 
(See \cite{tw1} for further discussion of these properties of the sine kernel.)\ \ 
\par
In this paper we show that there are completely analogous properties for a class of
kernels which arise when one rescales the Laguerre or Jacobi ensembles at the edge
of the spectrum. 
For the Laguerre ensemble the analogue of the sequence of
functions $\{\phi_{k}(x)\}$ in (\ref{KN}) is obtained by orthonormalizing the
sequence\[\{x^{k}\,x^{\al/2}\,e^{-x/2}\}\]over $(0,\iy)$\  (here $\al>-1$), 
whereas for Jacobi one orthonormalizes \[\{x^{k}(1-x)^{\al/2}(1+x)^{\beta/2}\}\]
over $(-1,1)$. (Here $\al,\beta\,>\,-1$.) In the Laguerre ensemble of (positive)
hermitian $N\times N$ matrices the eigenvalue density satisfies
\cite{bronk,nagao_wadati}, for a fixed $x<1$,
\[R_{1}(4Nx)\sim\frac{1}{2\pi}\sqrt{\frac{1-x}{x}}.\]
This limiting law is to be contrasted with the well-known Wigner semi-circle
law in the GUE. The new feature here is the ``hard edge'' for
$x\sim 0$.     
At  this  edge we have the scaling limit \cite{forrester}:
\[\lim_{N\ra\iy}\frac{1}{4N}K_{N}(\frac{x}{4N},\frac{y}{4N}) 
=\,\frac{ J_{\al}(\sqrt{x})\,\sqrt{y} J_{\al}'(\sqrt{y})\,-
\sqrt{x} J_{\al}'(\sqrt{x})\, J_{\al}(\sqrt{y})}{2(x-y)}\]
where $J_\al(z)$ is the Bessel function of order $\al$.
Both limits follow from the asymptotic formulas for the generalized Laguerre polynomials.
(Scaling in the bulk will just lead to the sine kernel and
scaling  at the ``soft edge,''
$x\sim 1$, will lead to  the Airy kernel.)\
The same 
kernel arises when scaling the Jacobi ensemble at $-1$ or $1$. (Recall
that in the Jacobi ensemble both $\pm 1$ are hard edges;  see e.g.~\cite{nagao_wadati}.)\ 
\par
For later convenience we introduce now a parameter $\la$ and define our ``Bessel 
kernel''  by
\begin{mathletters}\begin{eqnarray}
K(x,y):&=&\,\la\,\frac{J_{\al}(\sqrt{x})\,\sqrt{y}J_{\al}'(\sqrt{y})\,-
\sqrt{x}J_{\al}'(\sqrt{x})\,J_{\al}(\sqrt{y})}{2(x-y)}
\, ,\ \ \ (x\neq y)\label{Kdef}\\
&=&\, {\la\over 4}\left( J_{\al}(\sqrt{x})^2-J_{\al+1}(\sqrt{x})
 J_{\al-1}(\sqrt{x})
\right)\ \ \ (x=y).
\label{Kdiagdef}\end{eqnarray}\end{mathletters}
\par
Before stating our results, we mention that in the cases $\al=\mp\hf$ we have,
when $\la=1$,
\bq
2\sqrt{xy} \,K(x^{2},y^{2})\,=
\,\frac{\sin(x-y)}{\pi(x-y)}\,\pm\,\frac{\sin(x+y)}{\pi(x+y)},
\label{dyson_kernels}\eq
which are
  kernels which arise when taking scaling limits in the bulk of the spectrum 
for the Gaussian orthogonal and symplectic ensembles \cite{mehta_book}. In
particular, an asymptotic expansion we derive will generalize ones found by
Dyson \cite{dyson76} for the Fredholm determinants of these kernels.

We now state the results we have obtained.
\subsection{The System of Partial Differential Equations}
We set \bq J:= \bigcup_{j=1}^{m}(a_{2j-1},a_{2j}) \hspace{2em} (a_{j}\geq 0)
\label{Jform} \eq
and write $D(J;\la)$ for the Fredholm determinant of $K$ (the operator with kernel
$K(x,y)$) acting on $J$. If we think of this as a function of $a=(a_{1},\ldots,a_{2m})$
then
\bq d\log\,D(J;\la)\,=\,-\sum_{j=1}^{2m}(-1)^{j}\,R(a_{j},a_{j})\,da_{j}\label{diff}\eq
where $R(x,y)$ is the kernel of $K(I-K)^{-1}$.  We introduce the notations
\bq\phi(x):=\,\st{\la}\,J_{\al}(\st{x}),
\;\;\;\;\;\psi(x):=\,x\,\phi'(x)\label{phidef}\eq
and the quantities
\begin{eqnarray}
q_{j}:&=&\,(I-K)^{-1}\phi\,(a_{j}),~~p_{j}:=\,(I-K)^{-1}\psi\,(a_{j}),
\;\;\;\;(j=1,\ldots ,2m)\label{pqdef} \\
u:&=&\,(\phi,(I-K)^{-1}\phi),~~v:=\,(\phi,(I-K)^{-1}\psi),\label{uvdef} 
\end{eqnarray}
where the inner products refer to the domain $J$. The differential equations are
\begin{eqnarray}
\frac{\partial q_{j}}{\partial a_{k}}&=&(-1)^{k}\,
\frac{q_{j}p_{k}-p_{j}q_{k}}{a_{j}-a_{k}}  \,q_{k} \hspace{2em} (j\neq k)\label{qjk}\\
\noalign{\vskip3pt}
\frac{\partial p_{j}}{\partial a_{k}}&=&(-1)^{k}\,
\frac{q_{j}p_{k}-p_{j}q_{k}}{a_{j}-a_{k}}  \,p_{k} \hspace{2em} (j\neq k)\label{pjk}\\
\noalign{\vskip3pt}
a_{j}\frac{\pl q_{j}}{\pl a_{j}}&=&p_{j}\,+\,\frac{1}{4}\,q_{j}\,u\,
-\,\sum_{k\neq j}(-1)^{k}\,a_{k}\,\frac{q_{j}p_{k}-p_{j}q_{k}}{a_{j}-a_{k}}\,q_{k}
\label{qj} \\ \noalign{\vskip3pt}
a_{j}\,\frac{\pl p_{j}}{\pl a_{j}}&=&\frac{1}{4}(\al^{2}-a_{j}+2v)\,q_{j}
-\frac{1}{4}p_{j}\,u\,-\sum_{k\neq j}(-1)^{k}\,a_{k}\,\frac{q_{j}p_{k}-p_{j}q_{k}}
{a_{j}-a_{k}}\,p_{k} \label{pj}\\ \noalign{\vskip3pt}
\frac{\pl u}{\pl a_{j}}&=&(-1)^{j}\,q_{j}^{2} \label{uj}\\ \noalign{\vskip3pt} 
\frac{\pl v}{\pl a_{j}}&=&(-1)^{j}\,p_{j}\,q_{j}. \label{vj}\end{eqnarray}
Moreover the quantities $R(a_{j},a_{j})$ appearing in (\ref{diff}) are given by
\bq
a_{j}\,R(a_{j},a_{j})=\sum_{k\neq j}(-1)^{k}\,a_{k}\,\frac{(q_{j}p_{k}-
p_{j}q_{k})^{2}}{a_{j}-a_{k}}  
 +p_{j}^{2}\,-\,\frac{1}{4}\,(\al^{2}-a_{j}+2v)
q_{j}^{2}+\hf\,p_{j}\,q_{j}\,u. \label{Rjj}\eq 
These equations are quite similar to eqs.~(1.4)--(1.9) of \cite{tw2},
 as is their derivation.
\subsection{The ordinary differential equation}
For the special
 case $J=(0,s)$ the above equations can be used to show that $q(s;\la)$,
the quantity $q$ of the last section corresponding to the endpoint $s$, satisfies
\bq s(q^{2}-1)(s\,q')'\,=\,q(sq')^{2}+\qr(s-\al^{2})q+\qr sq^{3}(q^{2}-2)
\hspace{3em} ('=\frac{d}{ds}). \label{qeq}\eq
with boundary condition
\bq q(s;\lambda) \sim \fr{\sqrt{\lambda}}{2^\al\Gamma(1+\al)}\, s^{\al/2}\, ,
  \  \ \ \ s\rightarrow 0. \label{qeq_bc} \eq
This equation is reducible to a special case of the $P_{V}$
 differential equation;\footnotemark[3]
\footnotetext[3]{The Painlev{\'e} V differential equation is
\[ \fr{d^2y}{dx^2}=\left(\fr{1}{2y}+\fr{1}{y-1}\right) \left(\fr{dy}{dx}\right)^2
-\fr{1}{x}\fr{dy}{dx}+\fr{(y-1)^2}{x^2}\left(\al' y+\fr{\beta'}{y}\right) +
\fr{\gamma'  y}{x} + \fr{\delta' y(y+1)}{y-1}\]
where $\al'$, $\beta'$, $\gamma'$, and $\delta'$ are  constants.}
 explicitly, if $q(s)=(1+y(x))/(1-y(x))$ with
$s=x^2$, then $y(x)$ satisfies $P_V$
with $\alpha^\prime=-\beta^\prime=\alpha^2/8$,
$\gamma^\prime=0$ and $\delta^\prime=-2$.  (We have primed the usual
$P_V$ parameters to avoid confusion with the $\alpha$ in our kernel. We
mention that this special $P_V$ can be expressed algebraically in terms
of a third Painlev{\'e} transcendent and its first derivative \cite{gromak}.
We mention also that an argument can be given that (\ref{qeq}) 
must be reducible to one of the 50 canonical types of differential equations found
by Painlev{\'e}, without an explicit verification being necessary.  This
will be discussed at the end of section II B.)\ \
It is sometimes convenient to transform (\ref{qeq}) by making the substitution
\[ q(s)=\cos\psi(s),  \]
so that $\psi$ satisfies
\bq \psi^{\prime\prime} +{1\over s} \psi^\prime= {1\over 8s}\sin\left(
2\psi\right) -{\alpha^2\over 4s^2}\, {\cos\psi\over \sin^3\psi}\, .\label{psi_eq}\eq 
\par
The Fredholm determinant is expressible in terms of $q$ by the formula
\bq D(J;\la)\,=\,\exp\left(-\qr\int_{0}^{s}\log\frac{s}{t}\,q(t)^{2}\,
dt\right).
\label{Drep}\eq
Denoting by $R(s)$ minus the logarithmic derivative of $D(J;\la)$ with respect
to $s$, we have also the representation
\bq
R(s)=\qr\cos^2\psi(s) + s\left({d\psi\over ds}\right)^2 -
 {\alpha^2\over 4 s} \cot^2\psi(s) \, . \label{R_psi}
\eq
Furthermore, $R(s)$ itself satisfies a differential equation which in
the Jimbo-Miwa-Okamoto $\sigma$ notation for Painlev{\'e} III (see, in particular,
(3.13) in \cite{jimbo}) is
\bq
\left(s\sigma^{\prime\prime}\right)^2+\sigma^\prime\left(\sigma-s\sigma^\prime\right)
\left(4\sigma^\prime-1\right)-\alpha^2\left(\sigma^\prime\right)^2=0
\label{sigma_eq}\eq
where
$ \sigma(s)=s R(s)$; it
has small $s$ expansion
\begin{eqnarray}
\sigma(s;\lambda)&=&c_\al \, s^{1+\al}
 \left[1-\fr{1}{2(2+\al)} s +\fr{3+2\al}{16(1+\al)(2+\al)
(3+\al)} s^2 + \cdots\right]\nonumber \\ \noalign{\vskip3pt}
&&+ \fr{1}{1+\al}c_\al^2 \, s^{2+2\al}\left[1-
\fr{3+2\al}{2(2+\al)^2} s + \cdots\right]\nonumber \\ \noalign{\vskip3pt}
&& + \fr{1}{(1+\al)^2}c_\al^3\,  s^{3+
3\al}\left[1+\cdots\right]+\cdots
\end{eqnarray}
where
\[ c_\al = \fr{\lambda}{2^{2\al+2}} \fr{1}{\Gamma(1+\al)\Gamma(2+\al)}\, . \]
\par
We mention that in the special case $\al=0$
and $\lambda=1$  we have $D(J;1)=e^{-s/4}$,
$ q(s;1)=1$,
$\psi(s,1)=0$, and $\sigma(s,1)=s/4$
exactly \cite{edelman88,edelman91,forrester}.
\subsection{Asymptotics}
Again we take $J=(0,s)$ and consider asymptotics as $s\ra\iy$. From the random matrix
point of view the interesting quantities are 
\bq
E(n;s)\,:=\,\frac{(-1)^{n}}{n!}\,\frac{\pl^{n}}{\pl\la^{n}}\,
D(J;\la) \Bigr\vert_{\la=1}. \label{E_n} \eq
This is the probability that exactly $n$ eigenvalues lie in $J$. The asymptotics of
$E(0,s)=D(J;1)$ are obtained from (\ref{Drep}) using the asymptotics of $q(s;1)$
or equivalently $\sigma(s;1)$
obtained from (\ref{qeq})
or (\ref{sigma_eq}), respectively.
 (Our derivation is heuristic since as far as we are aware the corresponding
Painlev{\'e} connection problem has not been rigorously solved.)
We find that as $s\ra\iy$,
\begin{eqnarray}
E(0;s)&=&\tau_{\al}\,  {e^{-s/4+\al\st{s}}\over
s^{\al^2/4}} \label{E_asy}\\ \noalign{\vskip3pt}
&& \times\left(1+\frac{\al}{8}s^{-\hf}+\frac{9\al^{2}}{128}s^{-1}+(\frac{3\al}{128}+
\frac{51\al^{3}}{1024})s^{-\frac{3}{2}}+(\frac{75\al^{2}}{1024}
+\frac{1275\al^{4}}{32768})s^{-2}+\cdots\right),\nonumber
\end{eqnarray} 
where $\tau_{\al}$ is a constant which cannot be determined from the asymptotics of $q$
(or $\sigma$)
alone. However, as we mentioned above, when $\al=\mp\hf$ this expansion must agree with
those obtained from formulas (12.2.6) of \cite{mehta_book}
(see also (12.6.17)--(12.6.19) in \cite{mehta_book})
 after replacing $s$ by $\pi^{2}t^{2}$.
This leads to the conjecture
\[\tau_{\al}\,=\,\frac{G(1+\al)}{(2\pi)^{\al /2}} \]
where $G$ is the Barnes $G$-function \cite{barnes}.  This conjecture
is further supported by numerical work similar to that described
for the analogous conjecture in \cite{tw2}. 

As in \cite{tw2}, there are two approaches to the asymptotics of $E(n;s)$ for general $n$.
We use the notation
\bq r(n;s)\,:=\,\frac{E(n;s)}{E(0;s)}. \label{ratio} \eq

In the first approach (see also \cite{btw,tw1})
 one  differentiates (\ref{sigma_eq})
 successively with respect to $\la$.
Using the known asymptotics of $\sigma(s;1)$
and the differential equation (\ref{sigma_eq}) satisfied by $\sigma(s;\la)$ for
all $\la$,
one can find asymptotic expansions for
the quantities 
\[ \sigma_{n}(s)\,:=\,\frac{\pl^{n}\sigma}{\pl\la^{n}}\biggr\vert_{\la=1},\]
and these in turn can be used to find expansions for the $r(n;s)$. This approach is
inherently incomplete since yet another undetermined constant enters the picture. And
there are also computational problems since when one expresses the $r(n,s)$ in terms
of the $\sigma_{n}(s)$ a large amount of cancellation occurs, with the result that even
the first-order asymptotics of $r(n;s)$ are out of reach by this method when $n$
is large.
\par
The second approach uses the easily-established identity
\bq r(n;s)\,=\,\sum_{i_{1}<\ldots<i_{n}}\frac{\la_{i_{1}}\cdots\la_{i_{n}}}
{(1-\la_{i_{1}})\cdots(1-\la_{i_{n}})} \label{rexp}\eq
where $\la_{0}>\la_{1}>\cdots$\ are the eigenvalues of the integral operator $K$ with
$\la=1$ acting on $(0,s)$. It turns out that this operator, rescaled so that it acts on
$(0,1)$, commutes with the differential operator ${\cal L}$ defined by
\[ {\cal L}f(x)\,=\,(x(1-x)f'(x))'\,-\,(\frac{\al^{2}}{4x}+\frac{sx}{4})\,f(x),\]
with appropriate boundary conditions on $f$. Applying the WKB method to the equation,
and a simple relationship between the eigenvalues of $K$ (as functions of $s$) and its
eigenfunctions, we are able to derive the following asymptotic formula for the 
eigenvalues as $s\ra\iy$:
\bq1\,-\,\lambda_{i}\sim\frac{2\pi}{\Gamma(\al+i+1)\,i!}\,
s^{i+\frac{\al+1}{2}}\,e^{-2\sqrt{s}}\,2^{4i+2\al+2}.\eq
From this and (\ref{rexp}) we deduce
\bq r(n;s)\,\,\sim\left\{\prod_{k=0}^{n-1}\Gamma(\al+k+1)\,k!\right\}\pi^{-n}\,
2^{-n(2n+2\al+1)}\,s^{-\frac{n^{2}}{2}-\frac{\al}{2}n}\,e^{2n\st{s}}. \eq
\par
For the special case  $\al=0$, the quantity $r(1;s)$ can be expressed exactly
in terms of Bessel functions (see (\ref{bessel_r}) below).

\section{Differential Equations}
\subsection{Derivation of the system of equations}
We shall use two representations for our kernel. The first is just our definition
(\ref{Kdef}) using the notation (\ref{phidef}),
\bq K(x,y)=\frac{\phi(x)\psi(y)-\psi(x)\phi(y)}{x-y}. \label{Krep1} \eq
The second is the integral representation
\bq K(x,y)=\qr\int_{0}^{1}\phi(xt)\,\phi(yt)\,dt. \label{Krep2} \eq
This follows from the differentiation formula
\[z\,J_{\al}'(z)\,=\,\al\,J_{\al}(z)\,-\,z\,J_{\al+1}(z),\]
which gives the alternative representation
\[\la\,\frac{\st{x}\,J_{\al+1}(\st{x})\,J_{\al}(\st{y})\,-\,
J_{\al}(\st{x})\,\st{y}\,J_{\al+1}(\st{y})}{2\,(x-y)}\]
for $K(x,y)$, and the Christoffel-Darboux type formula (7.14.1(9)) of \cite{erdelyi}.

Our derivation will use, several times, the commutator identity
\bq [L,(I-K)^{-1}]\,=\,(I-K)^{-1}[L,K](I-K)^{-1}, \label{comid} \eq
which holds for arbitrary operators $K$ and $L$, and the differentiation formula
\bq\frac{d}{da}\,(I-K)^{-1}\,=\,(I-K)^{-1}\,\frac{dK}{da}\,(I-K)^{-1},\label{dform}\eq
which holds for an arbitrary operator depending smoothly on a parameter $a$.
We shall also use the notations
\[M\,=\,\mbox{multiplication by the independent variable},\hspace{2em}
D\,=\,\mbox{differentiation},\]
and a subscript on an operator indicates the variable on which it acts. 

It will be convenient to think of our operator $K$ as acting, not on $J$, but on 
$(0,\iy)$ and to have kernel
\[K(x,y)\,\chJ (y)\]
where $\chJ$ is the characteristic function of $J$. We continue to denote the
resolvent kernel of $K$ by $R(x,y)$ and note that it is smooth in $x$ but discontinuous
at $y=a_{j}$. The quantities $R(a_{j},a_{j})$ appearing in (\ref{diff}) are interpreted
to mean
\[\lim_{\stackrel{y\ra a_{j}}{y\in \sdown{J}}}R(a_{j},y),\]
and similarly for $p_{j}$ and $q_{j}$ in formulas (\ref{pqdef}). The definitions
(\ref{uvdef})  of $u$ and
$v$ must be modified to read
\bq u\,=\,(\phi\,\chJ,(I-K)^{-1}\phi),\hspace{2em}
v\,=\,(\phi\,\chJ,(I-K)^{-1}\psi),
\label{newuv}\eq where now the inner products are taken over $(0,\iy)$. Notice that 
since \[(I-K)^{-1}\xi\,=\,(I-K)^{-1}\xi\chJ\hspace{1em} \mbox{in}\ J\]
for any function $\xi$, this agrees with the original definitions (\ref{uvdef}) 
of $u$ and $v$.

We have, by (\ref{Krep2}),
\[((MD)_{x}+(MD)_{y})K(x,y)\,=\,\qr\int_{0}^{1}t\frac{\pl}{\pl t}
\left(\phi(xt)\phi(yt)\right)dt=\qr\phi(x)\phi(y)-K(x,y).\]
But it is easy to see that 
\bq [MD,L]\,\doteq\,((MD)_{x}+(MD)_{y}+I)\,L(x,y) \label{MDL} \eq
for any operator $L$ with kernel $L(x,y)$, where ``$\doteq$'' means ``has kernel''.
Taking $L(x,y)=K(x,y)\,\chJ(y)$ gives
\[[MD,K]\doteq\qr\,\phi(x)\,\phi(y)\,\chJ(y)\,-\,\sum(-1)^{k}a_{k}K(x,a_{k})\,
\delta(y-a_{k}).\]
(Recall the form (\ref{Jform}) of $J$.) It follows from this and (\ref{comid}) that
\bq[MD,(I-K)^{-1}]\,\doteq\,\qr\,Q(x)\,(I-K^{t})^{-1}\chJ\phi\,(y)
-\sum(-1)^{k}a_{k}R(x,a_{k})\,\rho(a_{k},y), \label{MDres} \eq
where $Q(x)$, and an analogous function $P(x)$, are defined by
\bq Q(x)\,:=\,(I-K)^{-1}\phi, \hspace{2em} P(x)\,:=\,(I-K)^{-1}\psi,\label{PQdef}\eq
where $\rho(x,y)=R(x,y)+\delta(x-y)$ is the distributional kernel of $(I-K)^{-1}$,
and where $K^{t}$ is the transpose of the operator $K$. (Note that $K$ takes smooth
functions to smooth functions while its transpose takes distributions to
distributions.) Observe that 
\[q_{j}\,=\,Q(a_{j}),\hspace{2em}p_{j}\,=\,P(a_{j}).\]

Next we consider commutators with $M$ and use the first representation (\ref{Krep1}) of
$K(x,y)$. We have immediately
\[ [M,K]\,=\,\left(\phi(x)\,\psi(y)\,-\,\psi(x)\,\phi(y)\right)\,\chJ(y),\]
and so, by (\ref{comid}) again,
\bq \left[M,(I-K)^{-1}\right]\,\doteq\,Q(x)\,(I-K^{t})^{-1}\psi\,\chJ(y)\,-\,
P(x)\,(I-K^{t})^{-1}\phi \chJ(y). \label{Mres} \eq
Notice that since
\[(I-K^{t})^{-1}\psi\chJ\,=\,(I-K)^{-1}\psi\,=\,P \hspace{2em} \mbox{on $J$},\]
and similarly for $\phi, Q$, we deduce 
\[R(x,y)\,=\,\frac{Q(x)\,P(y)-P(x)\,Q(y)}{x-y}\hspace{2em}(x,y\in J.) \]
In particular we have
\bq R(a_{j},a_{k})=\frac{q_{j}\,p_{k}-p_{j}\,q_{k}}{a_{j}-a_{k}}\hspace{3em}
(j\neq k) \label{Rjk} \eq
\bq R(x,x)\,=\,Q'(x)\,P(x)-P'(x)\,Q(x) \hspace{3em}(x\in J). \label{Rxx} \eq

In order to compute $R(a_{j},a_{j})$, and also the derivatives in (\ref{qj})
and (\ref{pj}), we must find $Q'(x)$ and $P'(x)$. We begin with the obvious
\[x\,Q'(x)=MD(I-K)^{-1}\phi\,(x)=(I-K)^{-1}MD\,\phi\,(x)+\left[MD,(I-K)^{-1}\right]\phi\,(x).\]
Using (\ref{MDres}), and recalling (\ref{phidef}) and (\ref{newuv}), we find that
\bq xQ'(x)=P(x)+\qr\,Q(x)\,u-\sum(-1)^{k}a_{k}R(x,a_{k})\,q_{k}.\label{Qder}\eq
Similarly, replacing $\phi$ by $\psi$ in this derivation gives
\bq xP'(x)=(I-K)^{-1}MD\,\psi\,(x)+\qr\,Q(x)\,v-\sum(-1)^{k}a_{k}R(x,a_{k})\,p_{k}.
\label{Pder1} \eq
To evaluate the first term on the right side we use the fact that $\phi$ satisfies
the differential equation
\bq x^{2}\,\phi''(x)+x\,\phi'(x)+\qr\,(x-\al^{2})\,\phi(x)=0,\label{besselDE}\eq
which may be rewritten $MD\,\psi\,(x)=\qr(\al^{2}-x)\,\phi$. Hence
\begin{eqnarray}
 (I-K)^{-1}MD\psi\,(x)&=&\fr{\al^{2}}{4}\,Q(x)-\qr(I-K)^{-1}M\phi\,(x) \nonumber \\
&=&\fr{\al^{2}}{4}\, Q(x)-\fr{x}{4}\, Q(x)+\qr\left[M,(I-K)^{-1}\right]\phi(x).
\label{Pder2}\end{eqnarray} 
But we find, using (\ref{Mres}), that
\[\left[M,(I-K)^{-1}\right]\phi\,(x)=Q(x)\,v-P(x)\,u,\]
and combining this with (\ref{Pder1}) and (\ref{Pder2}) gives
\bq x\,P'(x)=\qr(\al^{2}-x)\,Q(x)+\hf\,Q(x)\,v-\qr\,P(x)\,u
-\sum(-1)^{k}a_{k}R(x,a_{k})\,p_{k}. \label{Pder} \eq

It follows from (\ref{Rxx}), (\ref{Qder}) and (\ref{Pder}) that for $x\in J$
\[a_{j}\,R(a_{j},a_{j})=p_{j}^{2}-\qr(\al^{2}-a_{j}+2\,v)\,q_{j}^{2}+\hf p_{j}\,q_{j}
\,u+\sum_{k\neq j}a_{k}R(a_{j},a_{k})(q_{j}\,p_{k}-p_{j}\,q_{k}).\]
In view of (\ref{Rjk}) this is equation (\ref{Rjj}).

We now derive the differential equations (\ref{qjk})--(\ref{vj}). 
First, we have the easy fact that
\[ \fr{\pl}{\pl a_{k}}K\doteq(-1)^{k}K(x,a_{k})\,\delta(y-a_{k})\]
and so by (\ref{dform})
\bq \fr{\pl}{\pl a_{k}}(I-K)^{-1}\doteq(-1)^{k}R(x,a_{k})\,\rho(y,a_{k}).
\label{resder} \eq
At this point we use the notations $Q(x,a), P(x,a)$ for $P(x)$ and $Q(x)$ to remind
ourselves that they are functions of $a$ as well as $x$. We deduce immediately
from (\ref{resder}) and (\ref{PQdef}) that
\bq \fr{\pl}{\pl a_{k}}Q(x,a)=(-1)^{k}R(x,a_{k})q_{k},\;\;\fr{\pl}{\pl a_{k}}
P(x,a)=(-1)^{k}R(x,a_{k})p_{k}. \label{PQpls} \eq
Since $q_{j}=Q(a_{j},a)$ and $p_{j}=P(a_{j},a)$ this gives
\[\fr{\pl q_{j}}{\pl a_{k}}=(-1)^{k}R(a_{j},a_{k})q_{k},\hspace{1em}
\fr{\pl p_{j}}{\pl a_{k}}=(-1)^{k}R(a_{j},a_{k})p_{k},\hspace{2em}(j\neq k).\]
In view of
 (\ref{Rjk}) again, these are equations (\ref{qjk}) and (\ref{pjk}). Moreover
\[\fr{\pl q_{j}}{\pl a_{j}}=(\fr{\pl}{\pl x}+\fr{\pl}{\pl a_{j}})Q(x,a)\Bigr\vert_{x=a_{j}},
\hspace{1em}\fr{\pl p_{j}}{\pl a_{j}}=(\fr{\pl}{\pl x}+\fr{\pl}{\pl a_{j}})
P(x,a_{j})\Bigr\vert_{x=a_{j}}.\]
Equations (\ref{qj}) and (\ref{pj}) follow from this, (\ref{PQpls}), (\ref{Qder}), (\ref{Pder})
and (\ref{Rjk}). 

Finally, using the definition of $u$ in (\ref{newuv}), the fact
\[\fr{\pl}{\pl a_{j}}\chJ(y)=(-1)^{j}\delta(y-a_{j}),\]
and (\ref{resder}) we find that
\[\fr{\pl u}{\pl a_{j}}=(-1)^{j}\phi(a_{j})\,q_{j}+(-1)^{j}(\phi\chJ,R(\cdot,
a_{j}))\,q_{j}.\]
But
\[ (\phi\chJ,R(\cdot,a_{j}))=\int_{J}R(x,a_{j})\phi(x)\,dx=\int_{J}R(a_{j},x)
\phi(x)\,dx\]
since $R(x,y)=R(y,x)$ for $x,y\in J$. Since $R(y,x)=0$ for $x\not\in J$ the last
integral equals
\[\int_{0}^{\iy}R(a_{j},x)\phi(x)\,dx=q_{j}-\phi(a_{j}).\]
This gives (\ref{uj}), and (\ref{vj}) is completely analogous.

We end this section with two relationships (analogues of (2.18) and (2.19) of \cite{tw2})
which would allow us to express $u$ and $v$ in terms of the $q_{j}$ and $p_{j}$ if we
wished to do so. (They will also be needed in the next section.) These are
\bq 2v+\qr u^{2}+u=\sum_{j}(-1)^{j}a_{j}q_{j}^{2}, \label{int1} \eq
\bq u=\sum_{j}(-1)^{j}\left(4p_{j}^{2}-(\al^{2}-a_{j}+2v)q_{j}^{2}
+2p_{j}q_{j}u\right). \label{int2} \eq
To obtain the first of these observe that (\ref{qjk}) and (\ref{qj}) imply
\[ \left(\sum_{k}a_{k}\fr{\pl}{\pl a_{k}}\right)\,q_{j}=p_{j}+\qr q_{j}u,\]
while from (\ref{uj}) and (\ref{vj}),
\[\fr{\pl}{\pl a_{j}}(2v+\qr u^{2})=2(-1)^{j}q_{j}(p_{j}+\qr q_{j}u).\]
If we multiply both sides of the previous formula by $(-1)^{j}a_{j}q_j$
 and sum 
over $j$ what we obtain may be written
\[\left(\sum_{k}a_{k}\fr{\pl}{\pl a_{k}}\right)\,\left(\sum_{j}(-1)^{j}a_{j}q_{j}^{2}
  \right)-\sum_{k}(-1)^{k}a_{k}q_{k}^{2}
  =\left(\sum_{k}a_{k}\fr{\pl}{\pl a_{k}}\right)\,(2v+\qr u^{2}),\]
or equivalently
\[\left(\sum_{k}a_{k}\fr{\pl}{\pl a_{k}}\right)\,\left(\sum_{j}(-1)^{j}a_{j}q_{j}^{2}
\right)=\left(\sum_{k}a_{k}\fr{\pl}{\pl a_{k}}\right)\,(2v+\qr u^{2}+u).\]
It follows that the two sides of (\ref{int1}) differ by a function of 
$(a_{1},\ldots,a_{2m})$ which is invariant under scalar multiplication. Since, as is 
easily seen, both sides vanish when all $a_{j}=0$ their difference must vanish 
identically.

To deduce (\ref{int2}) we multiply (\ref{resder}) by $a_{k}$ and sum over $k$
and then add the result to (\ref{MDres}), recalling (\ref{MDL}), to obtain
\[\left( x\fr{\pl}{\pl x}+y\fr{\pl}{\pl y}+I+\sum_{k}a_{k}\fr{\pl}{\pl a_{k}}\right)
R(x,y)=\qr\,Q(x)\,Q(y)\]
for $x,y\in J$. This gives
\[ \left(\sum a_{k}\fr{\pl}{\pl a_{k}}\right)a_{j}R(a_{j},a_{j})=\qr a_{j}
q_{j}^{2}=\qr a_{j}(-1)^{j}\fr{\pl u}{\pl a_{j}}.\]
If we multiply both sides of this by $(-1)^{j}$ and sum over $j$ we deduce, by an 
argument similar to one just used, that
\bq \sum_{j}(-1)^j a_{j}R(a_{j},a_{j})=\qr u.\label{Ruform} \eq
Substituting for $a_{j}R(a_{j},a_{j})$ here the right side of (\ref{Rjj}) we see that the
resulting double sum vanishes, and (\ref{int2}) results.
\subsection{The ordinary differential equation}
In this section we specialize to the case $J=(0,s)$ and derive (among other things)
the differential equation (\ref{qeq}) and the representation (\ref{Drep}). 
In the notation of the
last section $m=1, a_{1}=0, a_{2}=s$. We shall write $q(s), p(s), R(s)$ for
$q_{2}, p_{2}, R(s,s)$, respectively. Equations (\ref{qj})--(\ref{vj}) become \begin{eqnarray}
s\,q'&=&p+\qr\,q\,u \label{qder} \\
s\,p'&=&\qr\,(\al^{2}-s)\,q+\hf\,q\,v-\qr\,p\,u \label{pder}\\
u'&=&q^{2} \label{uder}\\  v'&=&p\,q. \label{vder} \end{eqnarray}

It is immediate from (\ref{Ruform}) and (\ref{uder}) that
\bq (s\,R(s))'=\qr\,q(s)^{2},\label{Rq_eq}\eq
and since 
\[ \fr{d}{ds}\log D(J;\la)=-R(s)\]
(see (\ref{diff})), we obtain the representation (\ref{Drep}).
\par
To obtain the differential equation (\ref{qeq}) we apply $s\fr{d}{ds}$ to both sides of
(\ref{qder}) and use (\ref{qder}), (\ref{pder}) and (\ref{uder}). What results is
\bq s\,(s\,q')'=\qr(\al^{2}-s)q+\fr{1}{16}(u^{2}+8v)q+\qr s\,q^{3}.\label{ODE1} \eq
But (\ref{int1}) in this case is
\[u^{2}+8v=4s\,q^{2}-4u,\]
and so the above can be written
\bq s\,(s\,q')'=\qr(\al^{2}-s)q-\qr u\,q+\hf s\,q^{3}. \label{ODE2} \eq
Next, we square both sides of (\ref{qder}) and use (\ref{int2}), which now says
\[u=4p^{2}-(\al^{2}-s+2v)q^{2}+2p\,q\,u,\]
and find that
\[(s\,q')^{2}=\qr u+\qr(\al^{2}-s)q^{2}+\fr{1}{16}q^{2}(u^{2}+8v).\]
Combining this with (\ref{ODE1}) gives
\[s\,q\,(s\,q')'=(s\,q')^{2}-\qr u+\qr s\,q^{4},\]
and combining {\em this} with (\ref{ODE2}) gives the desired equation (\ref{qeq}).
The boundary condition (\ref{qeq_bc}) follows from the Neumann expansion of
the defining expression (\ref{pqdef})  for $q$. 
\par
Using (\ref{psi_eq}) one easily verifies that
\begin{eqnarray*}
 R(s)& =& \qr \cos^2\psi + s \left({d\psi\over ds}\right)^2-{\al^2\over 4s}\csc^2\psi
+ {c\over s}\\
&=& \qr \cos^2\psi + s \left({d\psi\over ds}\right)^2-{\al^2\over 4s}\cot^2\psi
+(c-{\al^2\over 4}){1\over s} 
\end{eqnarray*} 
satisfies (\ref{Rq_eq}) where $c$ is a constant of integration. That this constant
is equal to $\al^2/4$ follows from the small $s$ expansion of $R(s)$.  
(Use the fact that for $s\rightarrow 0$,  $R(s)\sim K(s,s)$ and that,  as follows
from (\ref{Kdiagdef}),  there is no simple pole in $s$.)\ \
Equation (\ref{sigma_eq}) follows from (\ref{R_psi}) and (\ref{psi_eq}).
\par
Here is the argument why (\ref{qeq}) must be reducible to some Painlev{\'e} equation
(or one of the other simpler differential equations on Painlev{\'e}'s list).
The derivation of (\ref{qeq}) used only the facts that the Bessel kernel
had both forms (\ref{Krep1}) and (\ref{Krep2}) and that the function $\phi$
satisfied the differential equation (\ref{besselDE}).  (Of course $\psi$ in (\ref{Krep1})
must be defined as $MD\phi$.)\ \  This equation has a 2-complex-parameter
family solutions and this gives a 2-complex-parameter family of kernels defined
by (\ref{Krep1}).  They can be shown to satisfy (\ref{Krep2}).  We replace the
kernels $K(x,y)$ by $s K(sx,sy)$ and have them act on $(0,1)$ rather than
$(0,s)$.  These operators on $(0,1)$ depend analytically on the {\it complex\/}
variable $s$ (except for a branch point at $s=0$) and the corresponding $q(s)$
can have, aside from a branch point at $s=0$, only poles which occur at the values
of $s$ for which $\la =1$ is an eigenvalue of the operator. (The resolvent
of an analytic family of compact operators has a pole whenever $\la =1$ is an
eigenvalue.) \ \  Thus the general solution (i.e., 2-complex-parameter family of
solutions) of (\ref{qeq}) has only poles as moveable singularities.  Since the
equation is of the form $q''=$~~rational function of $q'$ and $q$, it must be
reducible to one of the Painlev{\'e} types.\par
We mention that this argument requires $\vert\al\vert<1$ since it is only then
that all solutions of (\ref{besselDE}) give compact, or even bounded, operators
on $L_2$.  For other $\alpha$ it may be that we just have to replace $L_2$ by
an appropriate space of distributions.
\subsection{$r(1;s)$ for $\al=0$}
If we set  $\al=0$ and make the change of variables
 $s=x^2$, the differential equation for $\psi$ (recall (\ref{psi_eq})) becomes
\bq \psi^{\prime\prime} + \fr{1}{x} \psi^\prime = \fr{1}{2} \sin(2\psi)
\label{psi_eq2} \eq
and we want the solutions holomorphic at the origin. The linearization of this
differential
equation is the modified Bessel equation and all solutions of the linear equation
are linear combinations of $I_0(x)$ and $K_0(x)$. Flaschka and Newell \cite{flaschka}
have shown, using methods of monodromy preserving deformations
and singular integral equations, that the general
2-parameter solution to   (\ref{psi_eq2}) can be viewed as a ``perturbation'' of
this linear solution. (Precisely, they derive a singular
integral equation whose  Neumann expansion in a particular limit 
gives $\psi(x)$---the first term in this expansion is a 
linear combination of Bessel functions.)\ \
The one-parameter family of solutions  to (\ref{psi_eq2}) 
that are   holomorphic at the origin has the representation \cite{flaschka} 
\[ \psi(x;\mu) = \mu \psi_1(x) + \fr{\mu^3}{3!} \psi_3(x) + \fr{\mu^5}{5!}
 \psi_5(x) + \cdots \]
where
$ \psi_1(x) = I_0(x)$,
and
$ \mu^2=1-\lambda$. Note that we are using the slightly confusing
notation  $\psi(x;\mu)$  to denote the function $\psi$ of (\ref{psi_eq})
after the change of variables $s=x^2$ and $\mu^2=1-\la$. 
Multiple integral representations
(obtained from a Neumann expansion) for the higher $\psi_j$'s
 can be easily derived from 
\cite{flaschka}.
\par
The resolvent kernel $R(s)$, $s=x^2$, is given  in this special case by
\begin{eqnarray*}
4 R(s) &=& \cos^2\psi(x) +\left({d\psi\over dx}\right)^2 \\
&=& 1 + (1-\la) \left((\psi_1^\prime)^2-\psi_1^2\right)+(1-\la)^2
\left( \psi_1^4-\psi_1\psi_3 + \psi_1^\prime \psi_3^\prime\right) + \cdots
\end{eqnarray*}
Thus (recall (\ref{E_n}) and (\ref{ratio}))
\[ r(1;s)=-\int_0^s R_1(t)\, dt = -2 \int_0^{\sqrt{s}} x R_1(x^2)\, dx \]
where
\[ R_1(t) = \fr{\partial}{\partial\la} R(t)\biggr\vert_{\la=1}\> . \]
Therefore
\begin{eqnarray}
r(1;s)&=& \fr{1}{2} \int_0^{\sqrt{s}} x\left(\psi_1^2(x)-(\psi_1^\prime(x))^2\right)\, dx
\nonumber \\
&=& \fr{1}{2} \int_0^{\sqrt{s}} x\left( I_0(x)^2 - I_1^2(x)\right)\, dx \nonumber \\
&=& \fr{s}{2}\left( I_0^2(\sqrt{s})-\fr{1}{\sqrt{s}}\, I_0(\sqrt{s})I_1(\sqrt{s})
- I_1^2(\sqrt{s})\right)\, . \label{bessel_r}
\end{eqnarray}
The last equality follows from (5.542) of \cite{GR}.  
\par
We point out the curious fact that
(after letting $\psi\rightarrow i\psi$)  the same differential equation
(\ref{psi_eq2}) and
closely related  $\tau$-function arise in the 2D Ising model
\cite{wmtb,mtw,smj} {\it except\/} that 
here the boundary condition is  $\psi(x)\sim \mu  K_0(x)$ as
$x\rightarrow\infty$.
\bigbreak\bigbreak\bigbreak 
\section{Asymptotics}
\subsection{Asymptotics of the $\sigma$-equation}
In the case of the finite $N$ ensemble and $\al=0$, 
Edelman \cite{edelman88,edelman91} and Forrester \cite{forrester}
(by a direct evaluation of the integrals defining the probability $E_N(0;s)$)
has shown that $E(0;s)$
is exactly equal to $e^{-s/4}$.  From this it follows,   for $\alpha=0$
and $\lambda=1$,  that  $\sigma(s;1)=s/4$.
For general $\alpha$ and $\lambda=1$
 it is therefore reasonable to assume an asymptotic expansion
of the form:
\[ \sigma(s;1) =
c_1 s + c_2 s^{1/2} + c_3 + c_4 s^{-1/2} + \cdots \, , \ \ \  s\rightarrow\infty. \]
Substituting this into the differential equation (\ref{sigma_eq}) results in equations
that uniquely determine the coefficients $c_j$ {\it once\/} a choice 
in the square root $\sqrt{\al^2}$ is made.  Since for $\alpha=\mp\fr{1}{2}$ 
 our asymptotic
expansion of $E(0;s)$ must agree with those of Dyson (recall (\ref{dyson_kernels})),
 we
see that we must choose the square root $-\al$ $-\al$. A calculation then gives
\begin{eqnarray}
\sigma(s;1)&=&\fr{s}{4} -\fr{\al}{2}s^{1/2} + \fr{\al^2}{4} + \fr{\al}{16} s^{-1/2}+
\fr{\al^2}{16} s^{-1} + \fr{\al}{256}(16\al^2+9) s^{-3/2} 
 + \fr{\al^2}{64} (4\al^2+9)
s^{-2} \nonumber \\ \noalign{\vskip3pt}
&&\quad + \fr{\al}{2048}(128\al^4+720\al^2+225) s^{-5/2}+ \cdots \ \ \  s\rightarrow\infty ,
\label{sigma_asy}
\end{eqnarray}
from which  (\ref{E_asy}) follows.
\par 
\subsection{Asymptotics via the commuting differential operator}
Throughout this section we take $\la=1$. The operator $K$, when rescaled to act on 
$(0,1)$ instead of $(0,s)$, has kernel $s\,K(sx,sy)$. By (\ref{Krep2}) this is
equal to
\[ {s \over 4}\int_{0}^{1}\phi(sxt)\,\phi(syt)\,dt,\]
and so $K$ (rescaled, as it will be throughout this section), is the square of the
operator on $(0,1)$ with kernel
\[J(x,y)={\st{s}\over 2}\,\phi(sxy)={\st{s}\over 2}\,J_{\al}(\st{sxy}).\]
This function satisfies the differential equation
\bq x^{2}J_{xx}+xJ_{x}+(\frac{sxy}{4}-\frac{\alpha^{2}}{4})J=0.\label{W1} \eq
The operator $J$ will commute with a differential operator
\[{\cal L}= \frac{d}{dx}\alpha(x)\frac{d}{dx}+\beta(x)\]
if $\alpha(0)=\alpha(1)=0$ and if
\[\alpha(y)J_{yy}+\alpha'(y)J_{y}+\beta(y)J=\alpha(x)J_{xx}+\alpha'(x)J_{x}+\beta(x)J.\]
If we use (\ref{W1}) we see that this will be satisfied if 
\[ -\alpha(x)\left[x^{-1}J_{x}+(\frac{sy}{4x}-\frac{\alpha^{2}}{4x^{2}})\right]
+\alpha'(x)J_{x}+\beta(x)J \]
\[ =\mbox{the same expression with $x$ and $y$ interchanged.} \]
Equating the terms involving the first derivatives of $J$ gives
\[(\alpha'(x)-x^{-1}\alpha(x))J_{x}=(\alpha'(y)-y^{-1}\alpha(y))J_{y}.\]
But $x\,J_{x}=y\,J_{y}$, so the above will hold if
\[x^{-1}(\alpha'(x)-x^{-1}\alpha(x))=y^{-1}(\alpha'(y)-y^{-1}\alpha(y)).\]
This is satisfied if $\alpha(x)$ is a quadratic without constant term and, of
course, we choose  \[ \alpha(x)=x(1-x).\]
 What is required of $\beta$, then, is seen to be
\[\beta(x)-x(1-x)(\frac{sy}{4x}-\frac{\alpha^{2}}{4x^{2}})
=\beta(y)-y(1-y)(\frac{sx}{4y}-\frac{\alpha^{2}}{4y^{2}}),\]
which is satisfied by 
\[\beta(x)=-\frac{\al^{2}}{4x}-\frac{sx}{4}.\]
We write the differential equation,
 for which the eigenfunctions $f(x)$ are the eigenfunctions
of $J$, as
\bq (x(1-x)f'x))'+(\mu\sqrt{s}-\frac{\al^{2}}{4x}-\frac{sx}{4})f(x)=0.\label{W2}\eq
The boundary conditions are that $f(x)$ be bounded as $x\rightarrow1$ and that $f(x)$ 
be asymptotic to a constant times $x^{\al/2}$ as $x\rightarrow0.$\ The reason we
wrote the eigenvalues as we did is that for each $i$ the $\mu$ corresponding to the i'th
largest eigenvalue is bounded as $s\rightarrow\infty$. This is easily seen by an
oscillation argument. So we assume $i$ is fixed and proceed to find the asymptotics
of the corresponding eigenfunction $f(x)$ as $s\rightarrow\infty$. 
We assume it normalized so that 
\bq f(x)\sim x^{\al/2}\;\; \mbox{as}\; x\rightarrow 0.\label{W3}\eq

\subsubsection{The region $x\ll 1$}
The approximating equation is
\[f''(x)+\frac{1}{x}f'(x)+(\frac{\mu\sqrt{s}}{x}-\frac{\al^{2}}{4x^{2}}-
\frac{s}{4})f(x)=0.\]
The solution of this equation which satisfies (\ref{W3}) is
\[x^{\al/2}e^{-\sqrt{s}x/2}\Phi(\frac{1+\al}{2}-\mu,1+\al,\sqrt{s}x)\]
where $\Phi$ is the confluent hypergeometric function \cite{erdelyi}.  We deduce that
when $x\ll1$,
\bq f(x)\sim x^{\al/2}\,e^{-\sqrt{s}x/2}\,\Phi(\frac{1+\al}{2}-\mu,1+\al,\sqrt{s}x).
\label{W4} \eq

\subsubsection{The region      $x\ll 1, \protect\sqrt{s}   x\gg 1$}
\noindent{\em Case 1.} $\mu\neq i+\frac{1+\al}{2}\ (i=0,1,\ldots)$.
 Then from the known asymptotics
of $\Phi$ as its argument tends to $\infty$ (6.13(3) of \cite{erdelyi}) we deduce that
\bq f(x)\sim\frac{\Gamma(1+\al)}{\Gamma(\frac{1+\al}{2}-\mu)}\,
s^{-\frac{1}{2}(\frac{1+\al}{2}+\mu)}\,x^{-\frac{1}{2}-\mu}\,
e^{\sqrt{s}x/2}.\label{W5}\eq
{\em Case 2.} $\mu=i+\frac{1+\al}{2}$. Then
\bq \Phi(-i,1+\al,\sqrt{s}x)=\frac{\Gamma(\al+1)\,i!}{\Gamma(\al+i+1)}\,
L^{\al}_{i}(\sqrt{s}x)\label{W6}\eq             
where $L^{\al}_{i}$ is the generalized Laguerre polynomial
(6.9(36) of \cite{erdelyi}). So we
find that in this case
\bq f(x)\sim(-1)^{i}\frac{\Gamma(\al+1)}{\Gamma(\al+i+1)}\,s^{\frac{i}{2}}
\,x^{i+\frac{\al}{2}}\,e^{-\sqrt{s}x/2}.\label{W7}\eq
\subsubsection{The region  $\protect \sqrt{s}\; x\gg\, 1$,     $(1-x)\; s\gg\, 1$}

Here we use the standard WKB approximation for the solutions of a differential
equation $(p\,f')'+q\,f=0$\ given by $e^{\pm y}/(p\,q)^{\frac{1}{4}}$ where
$y=\int\sqrt{q/p}$. In the case of our equation (\ref{W2}), the range of validity of the 
approximation is as indicated in the heading of this section. To be definite, we take
\[y=-\int_{x}^{1}\sqrt{q(z)/p(z)}\,dz.\] It is easy to compute that for the 
range in question we have
\[y=-\sqrt{s}\,\sqrt{1-x}-\mu\,\log x+2\,\mu\log(1+\sqrt{1-x})+o(1).\]
Also, in this case $p\,q$ is asymptotically $-s/4$ times $x^{2}(1-x)$. Hence 
\[f(x)\sim\left(a(s)\,x^{-\mu}\,(1+\sqrt{1-x})^{2\mu}\,e^{-\sqrt{s}\sqrt{1-x}}\right.\]
\bq \left.+b(s)\,x^{\mu}\,(1+\sqrt{1-x})^{-2\mu}\,e^{\sqrt{s}\sqrt{1-x}}\right)
\label{W8}\eq
where $a(s)$ and $b(s)$ are constants depending on $s$.
\subsubsection{The region $x\rightarrow 1$}

Letting $y=1-x,f(x)=g(y)$ gives the approximating equation
\[ (y\,g(y)')'-\frac{s}{4}\,g(y)\,=\,0.\]
The general solution of this is a constant times $I_{0}(\sqrt{y})$ where $I_{0}$
is the modified Bessel function. Thus we deduce that as $x\rightarrow1$
\bq f(x)\sim\,c(s)\,I_{0}\left(\sqrt{s(1-x)}\right)\label{W9}\eq
for some c(s).
\subsubsection{Determination of $a(s)$, $b(s)$ and $c(s)$}

From the asymptotics of $I_{0}$ at infinity and (\ref{W9}) we deduce that
\[f(x)\sim\frac{c(s)}{\sqrt{2\pi}\,s^{\frac{1}{4}}}\frac{e^{\sqrt{s}\sqrt{1-x}}}
{(1-x)^{\frac{1}{4}}}\]
when $x\rightarrow1$ and $s(1-x)\rightarrow\infty.$ Comparing this with (\ref{W8}) 
shows that $a(s)=0$ and that 
\bq c(s)\sim\sqrt{2\pi}\,s^{\frac{1}{4}}\,b(s).\label{W10}\eq
And now comparing (\ref{W8}) with (\ref{W5}) and (\ref{W7}) in their overlapping range 
of validity we see that we must be in case 2 and that
\bq b(s)\sim(-1)^{i}\,\frac{\Gamma(\al+1)}{\Gamma(\al+i+1)}\,2^{2n+\al+1}\,
e^{-\sqrt{s}}.\label{W11}\eq

What we shall need from all this is, first, the asymptotics of $f(1)$. This follows
immediately from (\ref{W9}), (\ref{W10}), (\ref{W11}), and the fact that $I_{0}(0)=1$:
\bq f(1)\,\sim\,(-1)^{n}\,\sqrt{2\pi}\,\frac{\Gamma(\al+1)}{\Gamma(\al+i+1)}\,
s^{\frac{i}{2}+\frac{1}{4}}\,e^{-\sqrt{s}}2^{2n+\al+1}.\label{W12}\eq
We shall also need the asymptotics of $\int f(x)^{2}dx$. It follows from the
asymptotics we have derived that the main contribution to this integral comes from an
arbitrarily small neighborhood of $x=0$. It follows from (\ref{W4}), (\ref{W6}) and the 
fact  \[\int_{0}^{\infty}\,x^{\al}\,e^{-x}\,L_{i}^{\al}(x)^{2}\,dx\,=\,
\frac{\Gamma(\al+i+1)}{i!}\]
that  \[\int_{0}^{1}\,f(x)^{2}\,dx\sim
\frac{\Gamma(\al+1)^{2}\,i!}{\Gamma(\al+i+1}\,s^{-\frac{\al}{2}-\frac{1}{2}}.\]
We put these two relations together to get what we {\em really\/} want, which is
\bq \frac{f(1)^{2}}{\int_{0}^{1}\,f(x)^{2}\,dx}\sim\frac{2\pi}{\Gamma(\al+i+1)i!} \,
s^{i+\frac{\al}{2}+1}\,e^{-2\sqrt{s}}\,2^{4i+2\al+2}.\label{W13}\eq
\subsubsection{The asymptotics of $\lambda_{i}$} 

Since ${\cal L}$ commutes with our integral operator $K$, rescaled
to act on $(0,1)$, the set $\{f_{i}\}$ of eigenfunctions corresponding to the 
eigenvalues $\mu_{0}<\mu_{1}<\ldots$ of ${\cal L}$ is the set of
eigenfunctions of $K$ corresponding to its eigenvalues, in some order.\\ \mbox{}\\
{\bf Lemma 1}. {\em The eigenvalues of $K$ are simple}.\\ \mbox{} \\
\noindent {\bf Proof}. Since the eigenvalues of $K$ are the squares of the eigenvalues
of $J$ what we have to show is that if $f_{1}(x)$ and $f_{2}(x)$ are eigenfunctions
of ${\cal L}$ (they need not correspond to any particular $\mu$ here)
and if for some $\nu$ we have either
\[ \int_{0}^{1}J_{\al}(\sqrt{sxy})f_{i}(y)dy=\nu f_{i}(x)\;\;\;\;\;\;\;(i=1,2)\]
or
\[\int_{0}^{1}J_{\al}(\sqrt{sxy})f_{1}(y)dy=\nu f_{1}(x),\;\;
\int_{0}^{1}J_{\al}(\sqrt{sxy})f_{2}(y)dy=-\nu f_{2}(x) \]
then $f_{1}$ and $f_{2}$ are linearly dependent. We shall assume them normalized so
that they both have the value $1$ at $x=1$.

Notice first that $\nu\neq0$ for otherwise if we expand the $f_{1}$ equation near
$x=0$ we would find that \[\int_{0}^{1}x^{\frac{\al}{2}+k}f_{1}(x)dx=0,\;\; k=0,1,\ldots,\]
and $f_{1}$ would be identically 0.

Next we define \[H(x)=e^{-x/2}J_{\al}(\sqrt{s\,e^{-x}}),\;
g_{i}(x)=e^{-x/2}f_{i}(e^{-x}),\]
make the obvious variable changes, and find that our relations become
\bq\label{L1.1}\int_{0}^{\infty}H(x+y)\,g_{i}(y)\,dy=\nu\,g_{i}(x),\;\;\;\;(i=1,2),\eq
\bq\label{L1.2}\int_{0}^{\infty}H(x+y)\,g_{1}(y)\,dy=\nu\,g_{1}(x),\;\;
\int_{0}^{\infty}H(x+y)\,g_{2}(y)\,dy=-\nu g_{2}(x).\eq 

What comes now is almost identical to the proof of Lemma 1 of \cite{tw2}. Assuming first that
(\ref{L1.1}) holds, we differentiate twice this relation with $i=1$ and then integrate by
parts twice to obtain
\[\nu\,g_{1}(x)=-H'(x)+H(x)\,g_{1}'(0)+\int_{0}^{\infty}H(x+y)\,g_{1}(y)\,dy.\]
If we multiply both sides by  $g_{2}(x)$ and integrate, using (\ref{L1.1}) and its
differentiated version, we obtain (recall that $g_{i}(0)=1$)
\[\nu\int_{0}^{\infty}g_{1}''(x)\,g_{2}(x)\,dx=-\nu g_{2}'(0)+\nu g_{1}'(0)+
\nu\int_{0}^{\infty}g_{2}(y)\,g_{1}''(y)\,dy.\]
Thus, since $\nu\neq0$, we have $g_{1}'(0)=g_{2}'(0)$. Equivalently, $f_{1}'(1)
=f_{2}'(1)$. But since also $f_{1}(1)=f_{2}(1)$, it follows from equation (2)
that the corresponding eigenvalues $\mu$ must be the same, and so the eigenfunctions
are the same.

Next, assume (\ref{L1.2}) holds. Differentiating both sides of the first relation
once and integrating by parts give
\[\nu\,g_{1}'(x)=-H(x)-\int_{0}^{\infty}H(x+y)\,g_{1}'(y)\,dy.\]
Multiplying both sides of this by $g_{2}(x)$ and integrating, using the second part of
(\ref{L1.2}), we obtain
\[\nu \int_{0}^{\infty}g_{1}'(x)\,g_{2}(x)\,dx=-\nu+\nu\int_{0}^{\infty}g_{2}(y)
\,g_{1}'(y)\,dy,\]
contradicting $\nu\neq0$.\qed \\

Now that we know the eigenvalues $\lambda_{i}$ of $K$ are simple we can order them
so that $\lambda_{0}>\lambda_{1}>\ldots$\ . There is a permutation 
$\sigma$ of {\bf N} such that the eigenvalue corresponding to $f_{i}$ is 
$\lambda_{\sigma(i)}$. This permutation is independent of $s$ since the eigenvalues $\mu_{i}$
and $\lambda_{i}$ both vary continuously with $s$. The next lemma allows us to
compute $d\lambda_{i}/ds$ for each $i$.

\mbox{}

\noindent
{\bf Lemma 2}. {\em Let $\lambda$ be a simple eigenvalue of an integral operator on
$(0,1)$ with symmetric kernel of the form $s\,K(sx,sy)$ where $K(x,y)$ is smooth 
and symmetric. Let $f$ be a corresponding eigenfunction. Then}
\[\frac{d\lambda}{ds}\,=\,\frac{\lambda}{s}\frac{f(1)^{2}} 
{\int_{0}^{1}f(x)^{2}\,dx}.\]
{\bf Proof}. We may assume $f$ normalized so that $\int_{0}^{1}f(x)^{2}dx=1$. Then
according to Lemma 3 of \cite{tw2} we have
\[\frac{d\lambda}{ds}=\int_{0}^{1}\int_{0}^{1}\frac{\partial}{\partial s}(s\,K(sx,sy))\,
f(y)\,f(x)\,dy\,dx.\]
(This holds no matter what the form of the kernel.) In the case at hand the first factor
in the integrand equals
\[K(sx,sy)\,+\,(x\frac{\partial}{\partial x}\,+\,y\frac{\partial}{\partial y})K(sx,sy)\]
and we easily deduce from this that the integral itself equals
\[\frac{\lambda}{s}\left(\int_{0}^{1}f(x)^{2}dx\,+\,\int_{0}^{1}x\,f'(x)\,f(x)\,dx\,+\,
\int_{0}^{1}y\,f'(y)\,f(y)\,dy\right) \]
and integration by parts shows that the expression in parentheses equals $f(1)^{2}$.\qed
\par
{\noindent\bf Remark}.
 It is easily seen that the conclusion of the lemma holds when the kernel
has a mild singularity at $0$ such as our Besssel kernel has when $\al\,<\,0$.\\
\mbox{}\\
\noindent{\bf Lemma 3}. {\em For each $i$ we have $\lambda_{i}\,\rightarrow\,1$ as
$s\,\rightarrow\,\infty$.}\\\mbox{}\\
{\bf Proof}. The Hankel transform, when rescaled by the variable change 
$x\mapsto\sqrt{x}$, is the integral operator $H$ on $(0,\infty)$ with kernel
$\hf J_{\al}(\sqrt{xy})$ and so our operator $K$ on $(0,s)$ may be thought of as
$P_{s}\,H\,P_{s}\,H\,P_{s}$ where $P_{s}$ denotes the projection from $L_{2}
(0,\infty)$ to $L_{2}(0,s)$. Since, as is well-known, $H^{2}=I$, the minimax
characterization of the eigenvalues shows that for each $i$ the eigenvalue of
$P_{s}\,H\,P_{s}$ with i'th largest absolute value tends to $\pm\,1$ as $s\rightarrow
\infty$. Since our operator is the square of this one, the statement of the lemma
follows. \qed

We can now deduce the asymptotic formula (15) for the eigenvalues $\lambda_{i}$. 
We apply Lemma~2 to the eigenvalue $\lambda_{\sigma(i)}$ associated with the 
eigenfunction $f_{i}(x)$ of (\ref{W2}) and use (\ref{W13}) to deduce
\[\frac{d\,\log\lambda_{\sigma(i)}}{ds}\sim\frac{2\pi}{\Gamma(\al+i+1)\,i!}
\,s^{i+\frac{\al}{2}}\,e^{-2\sqrt{s}}\,2^{4i+2\al+2}.\]
Recalling Lemma 3 we see that we can integrate from $s$ to $\infty$ and we obtain
\[1\,-\,\lambda_{\sigma(i)}\sim\frac{2\pi}{\Gamma(\al+i+1)\,i!}\,
s^{i+\frac{\al+1}{2}}\,e^{-2\sqrt{s}}\,2^{4i+2\al+2}.\]

It remains to show that $\sigma(i)=i$ for all $i$. But it is clear from the above
formula that $i<j$ implies that $\lambda_{\sigma(i)}>\lambda_{\sigma(j)}$ for large
$s$ (and so for all $s$) and therefore $\sigma(i)<\sigma(j)$. Since $\sigma:{\bf N}
\rightarrow {\bf N}$ is onto, we must have $\sigma(i)=i$ for all $i$.

\acknowledgments
We wish to thank Professor F.~J.~Dyson  for
bringing  to our attention the notion of ``hard edge'' random matrix ensembles.  
This work was supported in part by the National Science
Foundation, DMS--9001794 and DMS--9216203, and this support is
gratefully acknowledged.

\end{document}